\documentclass[aps,prd,a4paper,onecolumn,amsmath,showpacs,superscriptaddress,nofootinbib,preprintnumbers,notitlepage]{revtex4-1}
\usepackage{amssymb,amsmath}
\usepackage{setspace}
\usepackage{graphicx}
\usepackage{color}
\usepackage{dcolumn}
\usepackage{pgfplots}
\usepackage{booktabs}
\usepackage{multirow}
\usepackage{dcolumn}
\usepackage{amsmath}
\usepackage{mathtools}
\usepackage{amsfonts}
\usepackage{amssymb}
\usepackage{epstopdf}
\usepackage{bm}
\usepackage{siunitx}
\usepackage{braket}
\usepackage{enumitem}
\usepackage{soul}
\usepackage{ulem}
\usepackage{color}
\usepackage{transparent}
\usepackage{pifont}
\usepackage[font={small}]{caption}
\newcommand\rf[1]{(\ref{eq:#1})}
\newcommand\lab[1]{\label{eq:#1}}

\newcommand\br{\begin{eqnarray}}
\newcommand\er{\end{eqnarray}}
\newcommand\be{\begin{equation}}
\newcommand\ee{\end{equation}}

\newcommand\lb{\lbrack}
\newcommand\rb{\rbrack}

\newcommand\bc{\begin{center}}
\newcommand\ec{\end{center}}

\newcommand\partder[2]{\frac{{\partial {#1}}}{{\partial {#2}}}}
\newcommand\eps{\epsilon}
\newcommand\vareps{\varepsilon}

\newcommand\G{\Gamma}

\renewcommand\k{\kappa}
\renewcommand\l{\lambda}

\newcommand\m{\mu}
\newcommand\n{\nu}

\renewcommand\P{\Phi}
\newcommand\pa{\partial}

\newcommand{\ct}[1]{\cite{#1}}
\newcommand{\bib}[1]{\bibitem{#1}}

\newcommand\PRD[3]{\textsl{Phys. Rev.} \textbf{D#1} (#2) #3}

\newcommand\PLB[3]{\textsl{Phys. Lett.} \textbf{#1B} (#2) #3}
\newcommand\CQG[3]{\textsl{Class. Quantum Grav.} \textbf{#1} (#2) #3}

\newcommand\IJMPD[3]{\textsl{Int. J. Mod. Phys.} \textbf{D#1} (#2) #3}

\newcommand\MPLA[3]{\textsl{Mod. Phys. Lett.} \textbf{A#1} (#2) #3}

\pgfplotsset{compat=1.18}
\begin{document}
\doublespacing
\title {Holomorphic gravity and its regularization of Locally Signed Coordinate Invariance \footnote{awarded honorable mention in the Gravity Research Foundation 2024 Awards for Essays on Gravitation}}
\date{15.05.2024}
\author{Eduardo Guendelman}
\email{guendel@bgu.ac.il}
\affiliation{Department of Physics, Ben-Gurion University of the Negev, Beer-Sheva, Israel.\\}
\affiliation{Frankfurt Institute for Advanced Studies (FIAS),
Ruth-Moufang-Strasse 1, 60438 Frankfurt am Main, Germany.\\}
\affiliation{Bahamas Advanced Study Institute and Conferences, 
4A Ocean Heights, Hill View Circle, Stella Maris, Long Island, The Bahamas.
}
%\date{Received: %date / Accepted: %date}
%\begin{document}
%\maketitle
\begin{abstract}
We expect the final theory of gravity to have more symmetries than we suspect and our research points in this direction. To start with, standard general coordinate invariance  can be extended to complex holomorphic general coordinate transformations. This is possible by introducing a non Riemannian
Measure of integration (NRMI) and where we avoid the non holomorphic  standard 
 $\sqrt{-g}$ measure of integration. Second, locally signed coordinate transformations where the Jacobian changes sign locally but the Jacobian approaches one asymptotically should be symmetries of Nature. This is unlike globally signed transformations that produce a change of boundary conditions, like in the cases of global parity and global time reversal, which are not symmetries of Nature.   The holomorphic extension can regularize the regions of space time where the Jacobian changes sign.  Consequences for Quantum Gravity are discussed.
\end{abstract}
%\keywords{modified gravity theories, %non-Riemannian volume forms, 
%global Weyl-scale symmetry spontaneous breakdown, %flat regions of scalar potential,
%non-singular origin of the universe}
%\PACS{04.50.Kd, % Modified theories of gravity
%11.30.Qc, % Spontaneous and radiative symmetry breaking
%98.80.Bp, % Origin and formation of the Universe
%95.36.+x % Dark energy
%}
%%%%%%%%%%%%%%%%%%%%%%%%%%%%%%%%%%%%%%%%%%%%%%%%%%%%%%%%%%%%%
%%%%%%%%%%%%%%%%%%%%%%%%%%%%%%%%%%%%%%%%%%%%%%%%%%%%%%%%%%%%%
\maketitle
\section{Introduction}
\label{intro}
The development of quantum theories of gravity has lead to the notion  that the quantum gravity theory
produces a phenomenology that involves the breakdown of symmetries, like Lorentz symmetry, etc. For a review see \ct{Quantum:gravityphenomenology}. We would like to explore the opposite possibility, that a more complete theory of gravity should have more space time symmetries than the ones considered so far.

As we will see,  this can be done in the framework of the metric independent non Riemannian measures has been used  for the construction of modified gravity theories
Refs.\ct{TMT-orig-1}-\ct{TMT-orig-3} (see also 
Refs.\ct{TMT-recent-1-a}-\ct{TMT-recent-2}). In some instances we have included the standard measure as well, where the standard Riemannian integration measure  might also contain a Weyl-scale symmetry preserving $R^2$-term \ct{TMT-orig-3}. Some applications have been: (i) $D=4$-dimensional models of gravity and matter fields containing  the new measure of integration appear to be promising candidates for resolution  of the dark energy and dark matter problems, the fifth force problem, and a natural mechanism for spontaneous breakdown of global Weyl-scale symmetry \ct{TMT-orig-1}-\ct{TMT-recent-2}, (ii) to study in Ref.\ct{susy-break} modified supergravity models with an alternative non-Riemannian volume form on the space-time manifold, (iii) resolving the Big Bang singularity by formulating models with a non singular emergent phase, which evolves into inflation and then transitions into a dark energy and dark matter phase  \cite{ourquintessence} - \cite{ourquintessencewithEDE},(iv) 
Gravity-Assisted Emergent Higgs Mechanism in the Post-Inflationary Epoch, \cite{Gravityassisted},  (v) To study reparametrization invariant theories of extended objects (strings and branes) based on employing a modified non-Riemannian 
world-sheet/world-volume integration measure \ct{mstring}, \ct{nishino-rajpoot}, which leads to a dynamically 
induced variable string/brane tension and to string models of non-abelian 
confinement, this leads to interesting consequences from the modified measures spectrum \ct{mstringspectrum}, and construction of new braneworld scenarios \ct{mstringbranes}.

Modifed Measures Theories have also been  discussed as effective theories for causal fermion theories \ct{MMT}. 

We will see here how in this framework one can construct a general coordinate invariant theory which has extended general coordinate transformations that includes also transformations with positive and  negative Jacobian, including locally alternating the sign in space time . This is possible by introducing a non Riemannian Measure of integration, which transforms according to the Jacobian of the coordinate transformation, not the absolute value of the  Jacobian of the coordinate transformation as it  is the case with $\sqrt{-g}$.

It is very important to notice that the signed general coordinate invariance is a particular case of the holomorphic general coordinate invariance in complex space, which can be used to regularize locally signed general coordinate invariance so the jacobian can go from positive to negative values without going through zero, using the complex plane to achieve this.
\section{ General Relativity and other  theories use a  Riemannian volume element that is not invariant under signed general coordinate transformations}
\label{GR}
The action of GR, and other theories that use the standard Riemannian volume element  $ d^4 x {\sqrt{-g}}$  is of the form,
\be
S = \int d^4 x {\sqrt{-g}} L
\lab{GRL}
\ee

where $L$ is a generally coordinate invariant lagrangian.
Now notice that under a general coordinate transformation, 
$$ d^4 x \rightarrow Jd^4 x $$  while 
$$ \sqrt{-g} \rightarrow  \mid J \mid ^{-1}\sqrt{-g}$$
where $J$ is the Jacobian of the transformation and $ \mid J \mid$ is the absolute value of the transformation. Therefore $d^4 x {\sqrt{-g}} \rightarrow  \frac{J}{ \mid J \mid} d^4 x {\sqrt{-g}} $, so invariance is achieved only for $J = \mid J \mid$, that is if $J>0$, that is signed general coordinate transformations are excluded.

One could argue that when taking the square root of the determinant of the metric one may choose the negative solution when it suits us, but this would be an arbitrary procedure if no specific rule is given to choose the positive or the negative root. We choose instead to declare that $ \sqrt{-g} $ is always positive and replace it in the measure by something else whose sign is well defined. 
\subsection{Invariance of the action with non invariant lagrangian density  (integrand) and compensating non invariant manifold of integration?}

If conditions are optimal, the non invariance of the lagrangian density  (integrand) , which includes the measure, in a signed coordinate transformation, could be compensated by the non invariance of the manifold of integration. For example, in a time reversal transformation, the integrand will change sign. But then we can  change the limits of integrations and the exchange of the limits of integrations will involve an additional exchange of signs that can compensate for the sign change in the integrand. 

Such transformations where the Jacobian of the transformation is negative all over space are for example a global time reversal transformation or in three spatial dimensions, a global parity transformation, which by the way are not symmetries of Nature, since they are broken by the weak interactions, see for example a modern textbook like  \ct{WEAKINTERACTION} for a review.

A very important issue is to separate signed general coordinate transformation which change boundary conditions that do not, since those transformations that change boundary conditions, like time reversal or parity, everywhere in space time do not appear to be symmetries of Nature, they are broken by the weak interactions, so they  seems not so relevant. More interesting would be signed general coordinate transformations in some regions of space time, but where  asymptotically,  at large values of the time and spacial coordinates,  the Jacobian becomes one. This possibility we will discuss as the most relevant. 
Such transformations that are locally signed are best formulated when non Riemannian measures are introduced which allows us to produce a  theory that allows holomorphic invariance, which is crucial in order to make sense of such transformations.

%Finally, we focus on local coordinate transformation that do not change boundary conditions. We %exclude for example $t \rightarrow -t $ which exchange also the limits of integration $(-\infty, %+ \infty) $ to  $(+\infty, - \infty) $ , while $dt \rightarrow -dt $, thus exchanging back the %limits of integration compensate the change in $dt$ and restores positivity. For localized %general coordinate transformations, that do not affect boundary conditions this possibility does %not exist .

%Indeed, considering a general coordinate transformation with negative jacobian $J$ all over $x$ space, we then have that, according to our previous finding, $d^4 x {\sqrt{-g}} \rightarrow  \frac{J}{ \mid J \mid} d^4 x {\sqrt{-g}} = - d^4 x {\sqrt{-g}} $, that then $$S \rightarrow -S$$,

%This change in sign in the action can be compensated by an antipodal transformation
%$$\phi \rightarrow \bar{\phi}$$ %$$\bar{\phi} \rightarrow \phi  $$   $$g_{\m\n} \rightarrow  \bar{g}_{\m\n}$$   $$ \bar{g}_{\m\n} \rightarrow g_{\m\n} $$ 

%This then produces an additional change in sign of the action that compensates the change in sign obtained from the signed coordinate transformation.

%This invariance of a signed general coordinate transformation, combined with the Linde´s antipodal transformation works as long as the jacobian is uniformly negative over all space.  

%%%%%%%%%%%%%%%%%%%%%%%%%%%%%%%%%%%%%%%%%%%%%%%%%%%%%%%%%%%%%
\section{Metric Independent Non-Riemannian Volume-Forms and Volume elements invariant under locally signed general coordinate transformations}
One can define a metric independent measure from a totally anti symmetric tensor gauge field, for example
\be
 \Phi (A) = \frac{1}{3!}\vareps^{\m\n\k\l} \pa_\m A_{\n\k\l} \quad ,
\lab{Phi}
\ee

Then, under a general coordinate transformation $$\Phi (A) \rightarrow J^{-1}\Phi (A). $$ Therefore $d^4 x \Phi(A)  \rightarrow  d^4 x \Phi(A) $, so invariance is achieved regardless of the sign of $J$.

\section{Theory using Metric Independent Non-Riemannian Volume-Forms}
\label{TMMT}

First we review our previous papers where we have considered  the  action 
of the general form involving two independent non-metric integration
measure densities generalizing the model analyzed in \ct{quintess} is given by 
\be
S = \int d^4 x\,\P_1 (A) \Bigl\lb R + L^{(1)} \Bigr\rb +  
\int d^4 x\,\P_2 (B) \Bigl\lb L^{(2)} + \eps R^2 + 
\frac{\P (H)}{\sqrt{-g}}\Bigr\rb \; .
\lab{TMMT1}
\ee

Here the following definitions  are used:

\begin{itemize}
\item
The quantities $\P_{1}(A)$ and $\P_2 (B)$ are two densities and these are  independent non-metric volume-forms defined in terms of field-strengths of two auxiliary 3-index antisymmetric
tensor gauge fields
\be
\P_1 (A) = \frac{1}{3!}\vareps^{\m\n\k\l} \pa_\m A_{\n\k\l} \quad ,\quad
\P_2 (B) = \frac{1}{3!}\vareps^{\m\n\k\l} \pa_\m B_{\n\k\l} \; .
\lab{Phi-1-2}
\ee
The density $\P (H)$ denotes  the dual field strength of a third auxiliary 3-index antisymmetric
tensor 
\be
\P (H) = \frac{1}{3!}\vareps^{\m\n\k\l} \pa_\m H_{\n\k\l} \; .
\lab{Phi-H}
\ee

\item
The scalar curvature $R = g^{\m\n} R_{\m\n}(\G)$ and the Ricci tensor $R_{\m\n}(\G)$ are defined in the first-order (Palatini) formalism, in which the affine
connection $\G^\m_{\n\l}$ is \textsl{a priori} independent of the metric $g_{\m\n}$.

\item
The two different Lagrangians $L^{(1,2)}$ correspond to two  matter field Lagrangians 
\end{itemize}
On the other hand, the variation of  \rf{TMMT1} w.r.t. auxiliary tensors 
$A_{\m\n\l}$, $B_{\m\n\l}$ and $H_{\m\n\l}$ becomes
\be
\pa_\m \Bigl\lb R + L^{(1)} \Bigr\rb = 0 \quad, \quad
\pa_\m \Bigl\lb L^{(2)} + \eps R^2 + \frac{\P (H)}{\sqrt{-g}}\Bigr\rb = 0 
\quad, \quad \pa_\m \Bigl(\frac{\P_2 (B)}{\sqrt{-g}}\Bigr) = 0 \; ,
\lab{A-B-H-eqs}
\ee
whose solutions are
\be
\frac{\P_2 (B)}{\sqrt{-g}} \equiv \chi_2 = {\rm const} \;\; ,\;\;
R + L^{(1)} = - M_1 = {\rm const} \;\; ,\;\; 
L^{(2)} + \eps R^2 + \frac{\P (H)}{\sqrt{-g}} = - M_2  = {\rm const} \; .
\lab{integr-const1}
\ee
Here the parameters $M_1$ and $M_2$ are arbitrary dimensionful and the quantity $\chi_2$ corresponds to an
arbitrary dimensionless integration constant. 

The resulting theory is called a Two Measure Theory, due to the presence of the two measures  $\P_1 (A) $ and $\P_2 (A) $. But for the purpose of this paper this is too general, since we want to restrict to a theory that will give us ordinary General Relativity, and we want to keep the general coordinate invariance under signed general coordinate invariance. 

For obtaining GR dynamics, we can restrict to one measure, so let us take 

$$ \P_1 (A)= \P_2 (B) = \Omega $$

 we can also express $ \Phi $ in terms of
four scalar fields
\be
\Omega = \frac{1}{3!}\vareps^{\m\n\k\l}\vareps^{abcd} \pa_\m\varphi_a \pa_\n\varphi_b \pa_\k\varphi_c \pa_\l\varphi_d \quad  
\lab{omega}
\ee
One has to point out that in the earlier formulations of modified measures theories we used this 4 scalar field representation for the measure, see for example\cite{TMT-orig-1} .
The mapping of the four scalars $\varphi_a$ to the coordinates $x^\mu$ may be topologically non trivial and this could be important in some discussions of the quantum theory. Finally, we have to correct the equation, 
\be
\frac{\P_2 (B)}{\sqrt{-g}} \equiv \chi_2 = {\rm const}  .
\lab{eq.tobe corrected}
\ee
for another equation the will be invariant under signed general coordinate invariant transformations, which will be 
\be
\frac{\Omega^2}{(-g)}\equiv \chi = K^2 =  {\rm const}  > 0 
\lab{corrected}
\ee
without loss of generality we define $K$ to be positive.
The resulting action that replaces \rf{Phi-1-2} is, 
\be
S = \int d^4 x\,\Omega \Bigl\lb R + L \Bigr\rb +  
\int d^4 x\,\Omega^2 \Bigl\lb 
\frac{\P (H)}{{(-g)}}\Bigr\rb \; .
\lab{simpleGCISIGNED}
\ee
the density $\P (H)$ remains defined eq. \rf{Phi-H}
so the integration obtained from the variation of the $H$ gauge field is  eq. \rf{corrected} now.
The solution of  eq. \rf{corrected} are
\be
\frac{\Omega}{ \sqrt{(-g)}} = \pm{ K}   .
\lab{solutions of corrected}
\ee
where the sign in \rf{solutions of corrected} will  be dynamically determined

Another possibility for a measure that would transform like the the Jacobian of the coordinate transformation, not the absolute value of the Jacobian,  would be the determinant of the vierbein. This will destroy however (up to a sign) the invariance of the theory under signed local Lorentz transformation of the vierbeins. that is Lorentz transformations with negative determinants, so, it is not a solution, rather we trade one asymmetry for another.

\section{Holomorphic general coordinate invariance invariance of the modified measure theory and regularization of signed coordinate transformations}
It is very useful to note that the action \rf{simpleGCISIGNED}, when extended to complex space time,has a much larger coordinate invariance appears, larger than signed general coordinate transformations. This is the group of holomorphic general coordinate transformations \cite{Holomorphic}, the reason being that in the action \rf{simpleGCISIGNED} the expression 
$\sqrt{-g}$, which is non holomorphic,  does not appear anymore now. This is enough to extend the general coordinate invariance in the complexified extension of the theory to holomorphic general coordinate invariance.

This allows us to regularize signed general coordinate transformations at some points in space time where the Jacobian would be zero if we have stayed in the real domain, while returning to the identity at very large times or very large values of the spatial coordinates. 

Indeed, the problem with considering signed general coordinate transformations, when restricting ourselves to a strictly real spacetime, is that the change from positive Jacobian to negative Jacobian will involve a singularity. Indeed, when  considering the Jacobian approach one at large values of the coordinates, but changing signs in the middle , when restricting to strictly real values, the Jacobian  must go from positive to negative values through zero or through infinity (approaching plus infinity then going to minus infinity). In either case, zero Jacobian or infinite Jacobian represent a singular transformations.

Using a holomorphic complex extension, we can regularize such sign changing Jacobian transformations, for example when just transforming one dimension, which we call $x$, like in,

$$x \rightarrow \bar{x} = \bar{x}  (x)$$ where,

\be
J =  
\frac{ \partial \bar{x}}{\partial x} = 
\tanh^2 (Ax) 
+ B(x + i\epsilon) exp(-\frac{x^2}{\Delta^2})
\lab{signed holomorphic transformation}
\ee
Where $A$, $B$ , $\Delta$ and  $\epsilon$ are real parameters and  we take $\epsilon$ very small, 
 we can see that, 

1. going through the real $x$ axis. we do not encounter any singularity,

2. As $x \rightarrow \pm{\infty} $, $J\rightarrow 1 $. So asymptotically we go to a trivial mapping, not affecting boundary conditions at infinity.

3. In the region close to $x=0$
the real part of J changes sign and no singularity in $J$ appears at $x=0$ due to the extension to complex space that allows going from positive to negative values of the Jacobian without singularities, that is due to the parameter $\epsilon$.
Since the  transformation \rf{signed holomorphic transformation} becomes trivial at asymptotic values,
the boundary conditions are not affected, and the manifold is invariant.

Boundary conditions may be  real, at least in the classical theory. All kind of features, like regularization of of symmetries, as we have seen above become more transparent by extending to the complexified theory and in the quantum theory, when tunneling and other effects are considered,  the complex space time becomes inevitable.

\section{A note on proper and improper signed coordinate transformations}
In gauge theories, gauge transformations that do not vanish fast enough at infinity,
are classified as improper gauge transformations, the transformations described in the previous sections, where we demand the Jacobian goes to one at infinity is a proper one, like in \rf{signed holomorphic transformation}. 

In contrast a signed transformation all over space is necessarily an improper one, since by definition the Jacobian does not return to one at infinity.

The analysis for the invariance of the action for the case of a signed transformation all over space involves changes in the limits of integration when the coordinates $x$ are considered. There may not be a need to change the limits of integration when the scalars used in the integration measure are considered as we will see in the next section.

\section{Only proper signed coordinate transformations have a chance to be symmetries of nature, possible connection to Bogoliubov transformations}
An important point to be made is that only proper signed coordinate transformations like in \rf{signed holomorphic transformation} have a chance to be symmetries of Nature. That is, only transformations with Jacobian approaching one at infinity. 

This of course excludes global Time Reversal and global  Parity transformations, which are of course known to be violated by the weak interactions. 

But then of course the proper signed coordinate transformations need that the Jacobian change sign in order to be negative in some regions of space and approach one at infinity. The change of sign cannot go strictly in the real space, because  in that case, the Jacobian changes sign locally either through zero or through infinity when the , so we invoke the holomorphic regularization, like in \rf{signed holomorphic transformation} requiring part of the transition to take place in the complex domain so as to avoid those singularities. 

Furthermore, it is important to note that the signed coordinate
transformation introduced here are transformation that
change the signature of the Jacobian and this is considered locally.
It resembles the Bogoliubov transformations which have been used in a non-trivial
curved space background, where they mixes the positive and
negative frequency components. See for example discussion on the possible connection of signed local general coordinate transformations and Bogoliubov transformations in \cite{Sayantan}.

\section{The invariant scalar integration manifold and invariant lagrangian density, another way to extend the space time manifold}
\label{invariantscalarintegrationmanifold}

Notice that using the volume element converts the the integration over coordinates in the action into integration over scalar fields, since $$\Phi d^4x = d\varphi_1 d\varphi_2 d\varphi_3 d\varphi_4$$

The integration manifold existing in the four scalar field manifold is in fact completely unaffected by any coordinate transformation taking place in the $x$ space. The lagrangian density is also a scalar not affected by any coordinate transformation, the theory formulated in this way does not require any boundary terms if the boundaries are for example formulated in the scalar field space.

In the case of \rf{simpleGCISIGNED} for example, $$S= \int  d\varphi_1 d\varphi_2 d\varphi_3 d\varphi_4 L ,$$
where
$$L=  \Bigl\lb R + L \Bigr\rb +  \,\Omega \Bigl\lb 
\frac{\P (H)}{{(-g)}}\Bigr\rb \; $$We have seen that complexifying the space can be a useful way to extend the space time manifold. Another way is to consider the measure scalars instead of the original coordinates, since the mapping between these two spaces may not be one to one.

In this respect, one issue that should be addressed is that of the gauge fixing in the $\varphi_a$ space. Indeed, we notice that the only thing where these fields appear in the equations of motion is $\Omega$, but this quantity is invariant under volume preserving diffeomorphisms of the fields $\varphi_a$, $\varphi^\prime_a = \varphi^\prime_a(\varphi_a)$  which satisfy

\be \lab{VOLPRESDIFF}
\epsilon_{a_1 a_2 a_3 a_4}\frac{\partial{\varphi^\prime}_{b_1}}{\partial\varphi_{a_1}}\frac{\partial{\varphi^\prime}_{b_2}}{\partial\varphi_{a_2}}\frac{\partial{\varphi^\prime}_{b_3}}{\partial\varphi_{a_3}}\frac{\partial{\varphi^\prime}_{b_4}}{\partial\varphi_{a_4}} = \epsilon_{b_1 b_2 b_3 b_4}
\ee

so the study of the best gauge for the $\varphi_a$  fields for further comparison with the  $x^\m$ space could be a very important subject. Of course when we say that the mapping between the $\varphi_a$ and the $x^\m$ spaces, we want to exclude multi valuedness due to volume preserving diffeomorphisms of the fields $\varphi_a$, if for example different signs for $\Omega$ are associated to the same point in  $x^\m$ space, it is clear that there are at least two points  in $\varphi_a$ space associated to one point in  $x^\m$ space, and these two points in  the $\varphi_a$ are not related through a volume preserving diff. 
\section{Gravitational Equations of motion}
\label{Einstein}

We start by considering the equation that results from the variation of the degrees of freedom that define the measure $\Omega$,  that is the scalar fields $\varphi_a$, these are,
\be
A^{\m}_a \pa_\m (R + L  +2 \Omega \frac{\P (H)}{{(-g)}}) = 0
\lab{EINSTEIN}
\ee
where 
\be
A^{\m a} =\frac{1}{3!}\vareps^{\m\n\k\l}\vareps^{abcd}  \pa_\n\varphi_b \pa_\k\varphi_c \pa_\l\varphi_d \quad  
\lab{AMATRIX}
\ee
Notice that the determinant of $A^{\m a}$ is proportional to $\Omega^3 $, so if the measure is not vanishing, the matrix $A^{\m a}$ is non singular and therefore $\pa_\m (R + L  +2 \Omega \frac{\P (H)}{{(-g)}}) = 0 $, 
so that,
\be
R + L  +2 \Omega \frac{\P (H)}{{(-g)}} = M = constant
\lab{M}
\ee

The variation with respect to the metric $g^{\m\n}$, we obtain.
\be
\Omega (R_{\m\n}+ \frac{\pa L}{\pa g^{\m\n} } )  +   g_{\m\n} \Omega^2 \frac{\P (H)}{{(-g)}} = 0
\lab{munu}
\ee
solving $\Omega \frac{\P (H)}{{(-g)}}$  from \rf{M} and inserting into \rf{munu}, we obtain,
\be
R_{\m\n} - \frac{1}{2}g_{\m\n} R  + \frac{1}{2} M g_{\m\n} + \frac{\pa L}{\pa g^{\m\n} } - \frac{1}{2}g_{\m\n} L = 0
\lab{EINSTEINLIKEEQ}
\ee
which gives exactly the form of Einstein equation with the canonical energy momentum defined from $L$

\be
T_{\m\n} = g_{\m\n} L - 2 \partder{}{g^{\m\n}} L \; .
\lab{EM-tensor}
\ee

The equations of motion of the connection (in the first order formalism) implies that the connection is the Levi Civita connection.  L can describe a scalar field with the potential and the term $\frac{1}{2} M $ can be interpreted as a shift of the scalar field potential by a constant or a floating contribution to the cosmological constant.
In the modified measure approach to this problem is best to consider the embedding space as the one defined by the four scalar fields $\varphi_a$ that define the measure $\Omega$ \rf{omega}. Therefore the most fundamental space is the  $\varphi_a$ , since only in this space we can formulate the full description and solution of the problem.

%%%%%%%%%%%%%%%%%%%%%%%
\section{Discussion and Conclusions}
In this essay we have seen that general coordinate invariance can be extended to locally signed transformations, where the Jacobian locally  changes sign . This is possible in models with modified measures of integration which are independent of the metric and avoid the square root ofthe determinant of the metric which is not holomorphic. The  holomorphic nature of the theory allows
to change the Jacobian of the transformation without any singularity, which is avoided through a consistent holomorphic regularization, where a small imaginary component is added to the Jacobian when we go from positive to negative values.

This small imaginary component in the Jacobian is reminiscent of the small imaginary component necessary to define the Feynman propagator so that positive and negative waves propagate forward or backwards in time. Recall also that a locally signed general coordinate transformation can indeed involve a local change in the direction of time, that could affect our notions of virtual pair creation which may become a pair annihilation process and vice versa.

Other interesting aspect is the necessary co existence of positive and negative measures in the theory, as expressed in eq. \rf{solutions of corrected}, so that according to this we could expect positive and negative measures of integration in the quantum gravity theory, which is a surprise, but recall that the appearance of positive and negative energies in Relativistic Quantum Mechanics was also a surprised, but Feynman provided us with a way to deal with this. In some sense this issue has been considered to some extent, For example,  Farhi et al \ct{FARHIGUVEN}, when considering the quantum nucleation of a baby universe, the authors are forced to consider a larger space than the original coordinate space, which would be the analog of our  scalars which can serve 
as integration variables as discussed in  section \ref{invariantscalarintegrationmanifold}, as Farhi et al \ct{FARHIGUVEN} had also have to consider that the additional spave covers multiple times the coordinate space, and with each covering, a sign for the measure could be positive or negative, so the calculations of Farhi et al \ct{FARHIGUVEN} can be interpreted in the context of the formalism of section \ref{invariantscalarintegrationmanifold}. 

Negative measures have been considered also by Linde in a non local model 
\cite{Lindemultiplication}
\be \lab{lindedoubleaction}
S = \int d^4 x d^4 y {\sqrt{-g(x)}} {\sqrt{-\bar{g}(y)}} (\frac{M_P^2}{16\pi} R(x) + L(\phi(x)) - \frac{M_P^2}{16\pi} R(y) - L((\bar{\phi}(y))
\ee
$R(x)$ is defined in terms of a $g_{\m\n}$ metric, while $ R(y)$  is defined in terms of a  $\bar{g}_{\m\n}$ metric and  
where $L(x)$ and $L(y)$ have exactly the same functional form with respect to their corresponding mirror fields, like for a scalar field $\phi(x)$, there will be a potential  $V(\phi(x))$, same with kinetic terms, etc. that are 
appearing in  $L(x)$ , while in $L(y)$  there will be corresponding field $\bar{\phi}(y)$ with a potential $V(\bar{\phi}(y))$, then in $L(y)$ the metric $\bar{g}_{\m\n}$ appears instead of 
the metric $g_{\m\n}$, then  the theory is obviously invariant under $V \rightarrow V + constant $ \cite{Lindemultiplication}. This non local coexistence of a spacetime and an antispacetime was shown by Linde to have remarkable properties concerning its behavior with respect to the cosmological constant problem. We can think of this in the context of our framework as  having regions where the measure of integration
can change sign, an effect that must take place at the same time as we double the space time, to realize  Linde´s ideas. 
Spaces with negative measures in gravity or in theories of extended objects we have called anti spaces, anti strings, anti branes, etc. Pair production of strings anti strings (formulated with a non Riemannian measure) can be considered \cite{stringsandantistrings}.

\section{Acknowledgements} I thank discussions with Mohammad Sami concerning possible connections between local signed general coordinate transformations and Bogoluibov transformations.
%%%%%%%%%%%%%%%%%%%%%%%%%%%%%%%%%%%%%%%%%%%%%%%%%%%%%%%%%%%%

\end{document}